\documentclass[prb,aps,amsmath,amssymb,superscriptaddress,reprint]{revtex4-2}
\usepackage{graphicx}
\usepackage{amssymb}
\usepackage{bm}

\begin{document}

\title{Accidental accuracy and formal consistency in $GW$+BSE:
Exact benchmarks and regime-dependent error cancellation}

\author{Michael O. Atambo}
\email{michael.atambo@tukenya.ac.ke}
\affiliation{Department of Physics, Earth and Environmental Science, Technical University of Kenya, Nairobi, Kenya}
\date{\today}

\begin{abstract}
Standard many-body perturbation theory for optical excitations combines the $GW$
approximation for quasiparticle energies with the Bethe-Salpeter equation (BSE)
for the electron-hole response. Formally, the BSE kernel must equal the
functional derivative of the self-energy, $K=\delta\Sigma/\delta G$. In routine
practice, however, the kernel is built from a statically screened direct
interaction and a bare exchange interaction, which breaks this consistency.
Using exact diagonalization of the extended Hubbard dimer, we construct a
controlled benchmark that isolates the price of this inconsistency. We show
that the internally inconsistent practical $GW$+BSE construction frequently
yields accurate optical gaps through accidental cancellation between an
underestimated $GW$ gap and an oversized bare-exchange kernel, and we map the
regions of parameter space where this accidental accuracy occurs. Upon switching on the nearest-neighbor interaction and binding the exciton,
the practical construction remains the accurate one throughout the
weak-binding regime, with cancellation optima at parts in $10^{3}$,
while the frozen-$W$ construction fails at every binding strength.
Beyond a binding threshold both static constructions deteriorate, showing
that deeply bound excitons require physics beyond any static kernel.
Accidental accuracy is therefore a property of the weak-binding regime, and
the systematic failure of the frozen-$W$ derivative kernel within the explored parameter space
traces to its overscreening of the exchange channel. The accuracy of
the practical kernel is an accidental cancellation of two independent errors,
an underestimated quasiparticle gap and an oversized exchange term, not a
hidden consistency. These results provide a controlled taxonomy of
approximation failure in $GW$+BSE and a precise diagnostic for when standard
workflows can be trusted.
\end{abstract}

\maketitle

\section{Introduction}

The combination of the $GW$ approximation~\cite{Hedin1965,Hedin1970} with the
Bethe-Salpeter equation (BSE) for the two-particle
response~\cite{Sham1966,Hanke1979,Hanke1980,Strinati1988,Onida2002,Martin2016} is the
standard route to optical spectra of real materials. The routine workflow is
schematically $G_0W_0 \rightarrow \{\varepsilon_n^{QP}\} \rightarrow$ static
BSE. This workflow is remarkably successful, and yet it mixes two
approximations that are formally unrelated: a dynamical quasiparticle
self-energy and an electron-hole kernel that is static and only partially
screened.

The formal structure of many-body perturbation theory prescribes a tighter
relation. In a conserving, $\Phi$-derivable
framework~\cite{Baym1961,Baym1962}, and in the equation-of-motion hierarchy
that closes the two-particle problem~\cite{Strinati1988,Onida2002}, the
electron-hole kernel is the functional derivative of the self-energy,
$K=\delta\Sigma/\delta G$. Practical implementations depart from this relation
in two controlled but uncontrolled-in-practice ways: the screened interaction
entering the direct channel is taken at zero frequency, and the exchange
channel is left bare. The consequences of dynamical screening and of vertex
structure beyond these simplifications have been investigated since the early
work of Strinati, Mattausch, and
Hanke~\cite{Strinati1980,Strinati1982}, who showed that retardation in the
electron-hole interaction can be essential in specific regimes. What has not
been quantified is a different question: what is the price of breaking the
self-energy/kernel consistency in the standard pipeline itself?

We ask precisely this question, in the spirit of a program that uses exact
constraints and exact benchmarks to expose when many-body approximations are
right, when they fail, and why~\cite{Atambo2025a,Atambo2025b}. The central
object of study is the comparison between three routes to the neutral
excitation spectrum: the exact route, the frozen-$W$ derivative route
$\Sigma^{GW}\rightarrow K^{GW}=\delta\Sigma^{GW}/\delta G|_W$, and the practical
route $\Sigma^{GW}\rightarrow K^{\rm prac}$. The question is not whether $GW$
is accurate, nor whether BSE is accurate, but whether an internally
inconsistent pair $(\Sigma,K)$ can nevertheless produce the correct excitation.

We address this question on the extended Hubbard dimer, solved exactly by
diagonalization in the $N=1,2,3$ sectors. The dimer is small enough that the
exact Green's function $G_{\rm exact}$, the exact self-energy
$\Sigma_{\rm exact}=G_0^{-1}-G_{\rm exact}^{-1}$, and the exact neutral
response $\chi_{\rm exact}$ are all available in closed Lehmann form, yet rich
enough to exhibit screening, correlation satellites, exchange splitting, and,
with a nearest-neighbor interaction $V$, a genuinely bound charge-transfer
exciton. The asymmetry $\Delta$ and the interactions $(U,V)$ serve as control
parameters for a two-dimensional anatomy of approximation error.

Our central result is a demonstration of regime-dependent error cancellation within the extended Hubbard dimer. In the weak-binding and resonance regimes the practical construction is accurate
for the wrong reason: the bare-exchange kernel overcompensates the
underestimated $GW$ gap, and the errors cancel, sometimes to parts in
$10^{3}$. The frozen-$W$ construction fails at every binding
strength because it overscreens the exchange channel. Upon increasing $V$ into the deep-binding regime,
both static constructions deteriorate as the exact gap outgrows any
static-kernel repair, locating the onset of retardation physics.
Inconsistency with the naive functional derivative is therefore benign where
binding is weak, and the failure of all static kernels where binding is
strong is sharp and computable.

The paper is organized as follows. Section~\ref{sec:exact} defines the model and the exact
benchmark objects, including the validation hierarchy and the
quasiparticle-versus-satellite criterion. Section~\ref{sec:gw} constructs the $G_0W_0$
baseline. Section~\ref{sec:kernels} states the kernel inconsistency and error decomposition at a
representative point. Section~\ref{sec:map} gives the parameter-space anatomy and the
regime crossing. Section~\ref{sec:discussion} discusses implications for first-principles
practice and for the design of repaired kernels.

\section{Exact many-body foundation}
\label{sec:exact}

We study the extended Hubbard dimer
\begin{equation}
H = -t\sum_{\sigma}\left(c^{\dagger}_{1\sigma}c_{2\sigma}+{\rm h.c.}\right)
+ U\sum_{i} n_{i\uparrow}n_{i\downarrow}
+ V\, n_1 n_2
+ \frac{\Delta}{2}\left(n_1-n_2\right),
\label{eq:H}
\end{equation}
with $t$ the hopping, $U$ the local interaction, $V$ the nearest-neighbor
interaction, and $\Delta$ a site asymmetry that lifts accidental degeneracies
and distinguishes local from charge-transfer character. All energies are
reported in units of $t$. The model is diagonalized exactly in each particle
number sector $N=0,\dots,4$, with dimensions $\binom{4}{N}$.

The exact retarded one-particle Green's function follows from the Lehmann
representation,
\begin{equation}
G_{ij}(\omega)=\sum_m
\frac{\langle\Psi_0|c_i|m\rangle\langle m|c_j^{\dagger}|\Psi_0\rangle}
{\omega-(E_m^{N+1}-E_0^{N})+i\eta}
+\sum_m
\frac{\langle\Psi_0|c_j^{\dagger}|m\rangle\langle m|c_i|\Psi_0\rangle}
{\omega-(E_0^{N}-E_m^{N-1})+i\eta},
\label{eq:G}
\end{equation}
with $\omega$ measured relative to the chemical potential
$\mu=\tfrac12[(E_0^{N+1}-E_0^{N})+(E_0^{N}-E_0^{N-1})]$. The exact self-energy
is then defined unambiguously through Dyson's equation,
\begin{equation}
\Sigma_{\rm exact}(\omega)=G_0^{-1}(\omega)-G_{\rm exact}^{-1}(\omega),
\label{eq:Sigma}
\end{equation}
where $G_0$ is the retarded Green's function of the same one-body part
$(t,\Delta)$ at the same filling. This choice of reference is essential: any
other $G_0$ would fold a redefinition of the non-interacting problem into
$\Sigma$ and corrupt the benchmark.

The exact neutral spectrum is obtained from the dipole (site-polarization)
response, with $d=n_1-n_2$,
\begin{equation}
\chi_{dd}(\omega)=\sum_{S}\frac{f_S}{\omega-\Omega_S+i\eta}
-\frac{f_S}{\omega+\Omega_S+i\eta},\qquad
f_S=|\langle 0|d|S\rangle|^2,
\label{eq:chi}
\end{equation}
where $\Omega_S=E_S^{N}-E_0^{N}$ are the exact neutral excitation energies.
Each eigenstate is classified by the spin Casimir
$S^2=S_-S_++S_z(S_z+1)$ evaluated on the degenerate multiplets, so that
singlets and triplets are identified without ambiguity, and the dipole
selection rule (the dipole is a spin scalar and couples only to singlets) is
verified numerically rather than assumed.

The implementation is certified by a hierarchy of independent tests: Hilbert
space dimensions, Hermiticity, the non-interacting and atomic limits,
particle-hole symmetry (which pins $\mu=U/2+V$ at half filling for any
$\Delta$), the spectral sum rule $\int A(\omega)d\omega=2$ per spin, and the
static-polarizability sum rule
\begin{equation}
\sum_S \frac{f_S}{\Omega_S}=-2\,\frac{\partial^2 E_0}{\partial\Delta^2},
\label{eq:sumrule}
\end{equation}
where the right-hand side is evaluated by finite differences of the
ground-state energy alone. Equation~\eqref{eq:sumrule} is satisfied to six
digits.

A further criterion is required before the language of satellites may be
used. Tracking the Lehmann weights of the addition and removal poles as a
function of $U$, we define the quasiparticle branch as the pole whose weight
flows to unity as $U\rightarrow 0$, and a correlation-induced satellite as a
pole whose weight vanishes in that limit. At the representative point
$(U,V,\Delta)=(4,0,0.5)$ the quasiparticle poles at $\pm1.805\,t$ carry weight
$Z=0.848$ and the shake-up poles at $\pm3.867\,t$ carry weight $0.152$.

At $V=0$ the lowest dipole-active singlet lies at $\Omega_1=4.636\,t$, above
the quasiparticle gap $E_{\rm gap}=3.611\,t$: the neutral excitation is a
charge-transfer resonance, not a bound exciton. The nearest-neighbor
interaction $V$ supplies the bare electron-hole attraction that binds the
pair, and is therefore the control parameter of the bound-exciton regime.

\section{The $G_0W_0$ baseline}
\label{sec:gw}

The screened interaction is constructed in the site-density basis from the
non-interacting bubble $\chi^0$ of the same consistent $G_0$ used in
Eq.~\eqref{eq:Sigma}, with bare interaction matrix
$v_{ij}=U\delta_{ij}+V(1-\delta_{ij})$:
\begin{equation}
\varepsilon(\omega)=1-v\chi^0(\omega),\qquad
W(\omega)=\varepsilon^{-1}(\omega)\,v .
\end{equation}
For the dimer, $\det\varepsilon$ has a single positive root, the screened
charge mode $\Omega_p$, and the correlation part of the self-energy is
obtained in closed form,
\begin{equation}
\Sigma^{c}_{ij}(\omega)=\sum_k P^k_{ij}\,R_{ji}
\left[\frac{f_k}{\omega-\varepsilon_k-\Omega_p-i\eta}
+\frac{1-f_k}{\omega-\varepsilon_k-\Omega_p+i\eta}\right],
\label{eq:Sigc}
\end{equation}
where $P^k$ are the one-body projectors, $R$ is the residue of $W$ at
$\Omega_p$, and the causal structure follows the occupation factors. The full
self-energy is $\Sigma^{GW}=\Sigma_H+\Sigma_x+\Sigma^c$, with Hartree and Fock
terms built from the $G_0$ densities; at $V=0$ and half filling the static
part vanishes identically, so $GW$ is purely dynamical.

At $(U, V, \Delta) = (4, 0, 0.5)$ the $GW$ quasiparticle gap is $2.736\,t$ against the exact $3.611\,t$: RPA overscreens and the gap is underestimated by $0.875\,t$. The $GW$ satellite poles sit at $\varepsilon_k+\Omega_p\approx
\pm4.50\,t$, misplaced by $+0.63\,t$ relative to the exact shake-up poles. In
this minimal model the $GW$ satellite moreover carries vanishing residue in
the dressed propagator: the satellite structure is present in $\Sigma^{GW}$
but spectroscopically invisible in $G^{GW}$. The $GW$ baseline is therefore a
controlled, quantitatively wrong reference, which is precisely what the
consistency question requires.

\section{Kernel inconsistency and error cancellation}
\label{sec:kernels}

In the two-level Tamm-Dancoff approximation the neutral excitations built on
the $GW$ quasiparticles read
\begin{equation}
\Omega_S=E_{\rm gap}^{QP}+2K^x-K^d,\qquad
\Omega_T=E_{\rm gap}^{QP}-K^d,
\label{eq:bsesimple}
\end{equation}
with $K^x$ and $K^d$ the exchange and direct electron-hole kernel elements in
the transition basis. These elements are molecular integrals over the
occupied and unoccupied one-body orbitals $\psi_o$ and $\psi_u$ of the
reference Hamiltonian. With the transition density
$\rho^{\rm tr}_i=\psi_o(i)\psi_u(i)$ and the level densities
$\rho^{o}_i=\psi_o(i)^2$ and $\rho^{u}_i=\psi_u(i)^2$, the singlet kernel
combinations for the two constructions read
\begin{eqnarray}
2K^x_{\rm prac}-K^d &=& 2\sum_{ij} v_{ij}\,\rho^{\rm tr}_i\rho^{\rm tr}_j
-\sum_{ij} W_{ij}(0)\,\rho^{o}_i\rho^{u}_j,
\label{eq:kprac}\\
2K^x_{\rm cons}-K^d &=& 2\sum_{ij} W_{ij}(0)\,\rho^{\rm tr}_i\rho^{\rm tr}_j
-\sum_{ij} W_{ij}(0)\,\rho^{o}_i\rho^{u}_j .
\label{eq:kcons}
\end{eqnarray}
The construction $K^{\rm cons}$ is the frozen-$W$ functional derivative of $\Sigma^{GW}$, obtained by holding the screened interaction fixed; it is the object one obtains by
applying the consistency slogan $K=\delta\Sigma/\delta G$ without the vertex
term. Whether including the vertex term $iG(\delta W/\delta G)$ restores the
bare exchange, as continuum arguments suggest, is tested explicitly in
Appendix~\ref{app:vertex}. The direct term is common to both constructions.

At $(U,V,\Delta)=(4,0,0.5)$ the exact singlet is $\Omega_1=4.636\,t$. The
practical construction gives $4.542\,t$, an error of $-0.094\,t$, while the
frozen-$W$ construction gives $2.110\,t$, an error of $-2.527\,t$. The
triplet, which is exchange-free, is reproduced by both constructions at
$0.778\,t$ against the exact $0.836\,t$, an error of $-0.058\,t$. The singlet
result decomposes exactly, as summarized in Table~\ref{tab:decomp}: the $GW$
gap error is $-0.875\,t$; the practical kernel overestimates the exact
required kernel $\Omega_1-E_{\rm gap}=1.026\,t$ by $+0.780\,t$, and the sum is
$-0.094\,t$; the frozen-$W$ kernel errs by $-1.652\,t$, which added to the
gap error yields $-2.527\,t$. The practical construction is therefore
accurate through a Type-IV accidental cancellation, a right exciton produced
by wrong quasiparticles and a wrong kernel, while the frozen-$W$ construction
stacks two same-sign errors. The kernel inconsistency measure
$\mathcal{I}_K=|2K^x_{\rm prac}-2K^x_{\rm cons}|=2.432\,t$ is large, yet the
optical error of the inconsistent construction is small. At this point of the
phase diagram, inconsistency does not predict excitation error.

\begin{table}[b]
\caption{Error decomposition of the neutral excitation energies at
$(U,V,\Delta)=(4,0,0.5)$, in units of $t$. The practical construction is
accurate because its positive kernel error compensates the negative
quasiparticle gap error. The frozen-$W$ construction adds the two errors.
The triplet is exchange-free and therefore insensitive to the kernel choice.}
\label{tab:decomp}
\begin{ruledtabular}
\begin{tabular}{lcc}
Quantity & Value & Error \\
\hline
$E_{\rm gap}^{GW}$ & 2.736 & $-0.875$ \\
Exact kernel, $\Omega_1-E_{\rm gap}^{ex}$ & 1.026 & reference \\
$2K^x-K^d$, practical & 1.806 & $+0.780$ \\
$2K^x-K^d$, frozen-$W$ & $-0.626$ & $-1.652$ \\
$\Omega_S$, practical & 4.542 & $-0.094$ \\
$\Omega_S$, frozen-$W$ & 2.110 & $-2.526$ \\
$\Omega_T$, both kernels & 0.778 & $-0.058$ \\
$\Omega_1$, exact & 4.636 & exact
\end{tabular}
\end{ruledtabular}
\end{table}

\section{Parameter-space anatomy and the regime crossing}
\label{sec:map}

\begin{figure*}[t]
\includegraphics[width=\textwidth]{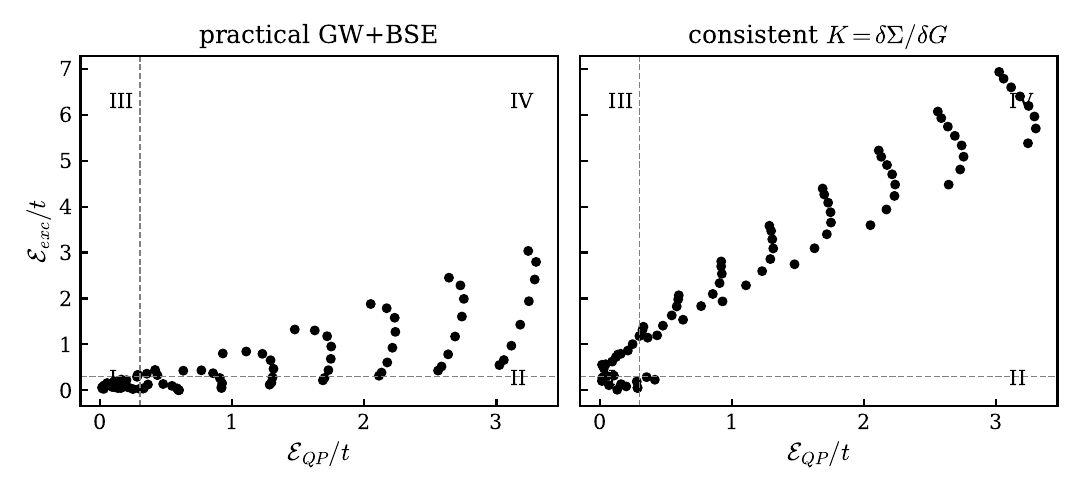}
\caption{Accuracy map over the $(U,\Delta)$ plane at $V=0$: joint
quasiparticle error $\mathcal{E}_{QP}$ and excitation error
$\mathcal{E}_{exc}$ for the practical (left) and frozen-$W$ (right)
constructions. The practical construction populates the wrong-quasiparticle,
right-exciton quadrant; the frozen-$W$ construction does not.}
\label{fig:map}
\end{figure*}

To convert the single-point error decomposition into a structural result we sweep the
parameter space. Figure~\ref{fig:map} reports the joint quasiparticle and
excitation errors over the $(U,\Delta)$ plane at $V=0$. The practical
construction populates a broad accidental-accuracy region in which
$\mathcal{E}_{QP}$ reaches $3.3\,t$ while $\mathcal{E}_{exc}$ remains below
$0.3\,t$: of 80 parameter points, 17 lie deep in this wrong-quasiparticle,
right-exciton region and 20 in the jointly accurate region. The frozen-$W$
construction empties the accidental region almost completely, retaining only 2
points, and its excitation error grows in proportion to the quasiparticle
error. Inconsistency is benign over a wide domain of the phase diagram, and
the frozen-$W$ construction is accurate only where $GW$ itself happens to be accurate.

Switching on $V$ at $(U,\Delta)=(4,0.5)$ binds the exciton: the optical gap
detaches below the quasiparticle gap for $V\gtrsim 1.0\,t$, reaching a
binding energy of $2.60\,t$ at $V=3\,t$. Figure~\ref{fig:crossing} shows the
corresponding excitation errors, evaluated with the analytic pole solver of
Appendix~\ref{app:poles}. The practical error is non-monotonic, with two
cancellation sweet spots at $V\simeq0.97\,t$ and $V\simeq1.80\,t$ where the
error falls to a few $\times10^{-3}\,t$, followed by a steady growth as the
exact gap increases with $V$ faster than the $GW$ gap. The frozen-$W$ error
remains of order $1$ to $2\,t$ at every binding strength. There is no regime
within the explored parameter space in which the frozen-$W$ derivative kernel is the accurate construction.

The physics is transparent in the kernel channels. In weak binding the
singlet is exchange-sensitive, and the bare exchange of the practical kernel
supplies the compensating repulsion that repairs the $GW$ gap deficit; the
frozen-$W$ kernel, which overscreens the exchange, has no such compensation.
In deep binding the situation changes qualitatively: the exact gap grows
with $V$ while the $GW$ gap is pinned by screening, and no static kernel can
reconcile the two. The deterioration of both constructions beyond
$V\simeq2\,t$ is therefore not a kernel-choice issue but a signature of
missing retardation physics, in line with the early analysis of Strinati,
Mattausch, and Hanke~\cite{Strinati1980,Strinati1982}.

\section{Discussion and outlook}
\label{sec:discussion}

Our results reframe the question of when the standard $GW$+BSE workflow can be
trusted. Conventional three-dimensional semiconductors, whose optical
excitations are weakly bound or resonant, may sit in a regime analogous to the weak-binding sector of the dimer, where
the practical construction benefits from error cancellation. This provides a possible mechanism for
its persistent empirical success despite formally inconsistent
inputs, though first-principles testing is required to confirm this mapping. Tightly bound excitons in low-dimensional materials, organic crystals,
and moir\'e systems may sit on the other side of the crossing, where the same
workflow loses its accidental protection. The crossing point and the sweet spot are in principle computable
diagnostics, and the kernel inconsistency measure $\mathcal{I}_K$ provides an
inexpensive indicator that could be evaluated alongside a standard BSE run as a proposed diagnostic.

Appendix~\ref{app:vertex} evaluates the vertex term $iG(\delta W/\delta G)$ explicitly in
the dimer by propagating a coherence-direction perturbation of the reference
Green's function through the full RPA chain. The vertex contribution does
not reduce the exchange channel to the bare interaction: the residual
relative to the continuum cancellation identity is of order $t$ across the
parameter range tested (Table~\ref{tab:vertex}). The practical kernel is therefore not the frozen-$W$
kernel in disguise, and the cancellation that makes it accurate is genuinely
accidental in the sense of Table~\ref{tab:decomp}: two independent accuracy errors, one in
the quasiparticle gap and one in the exchange channel, compensate. The
continuum cancellation cited in Refs.~\cite{Strinati1988,Onida2002} relies
on momentum integration and dynamical limits that a finite, static,
rank-one polarizability does not possess, and its failure here is a feature
of the minimal Hilbert space rather than a defect of the benchmark. The practical BSE workflow remains an internally inconsistent approximation that succeeds via error cancellation, not via a hidden formal consistency.

The systematic failure of the frozen-$W$ derivative kernel within the explored parameter space, and the
deterioration of the practical kernel beyond $V\simeq2\,t$, locate the
limits of the static picture. For deeply bound excitons the relevant energy
scales approach the plasmon frequency and dynamical screening becomes
essential; no static kernel, practical or frozen-$W$, can then reproduce the
exact spectrum, and repairing the BSE requires explicitly dynamical kernels
and retardation effects in the electron-hole
interaction~\cite{Strinati1980,Strinati1982}. The regime-dependent
error cancellation is thus refined to a two-part statement: bare exchange
compensating the gap error explains the accuracy of the
practical kernel in weak binding, and the breakdown of all static kernels
explains its failure in deep binding.

The role of retardation in the electron-hole interaction was established in
the foundational work of Strinati, Mattausch, and
Hanke~\cite{Strinati1980,Strinati1982}, who showed that dynamical screening
can be essential for excitonic and satellite structure. The present work is
complementary rather than overlapping: we do not compute dynamical
corrections; we ask how and why the static construction succeeds when it
succeeds. Using exact benchmarks we decompose the static error channel by
channel, map the domain of accidental accuracy, and show that the deep-binding failure of all
static kernels is the quantitative onset of the regime identified in
Ref.~\cite{Strinati1980}. Strinati located the regime; the exact cartography
of the static limit and its computable breakdown
are the contribution of this paper.

Several extensions define the forward program. Dynamical kernels, in the
spirit of the early work on retardation in the electron-hole
interaction~\cite{Strinati1980,Strinati1982}, will be needed to repair the
satellite sector, where $GW$ misplaces the shake-up poles and the static BSE
cannot respond. The relation between conservation laws and spectral accuracy,
and the construction of moment-based a priori diagnostics for kernel
reliability, connect this program to the broader theme of exact constraints in
many-body approximations~\cite{Baym1961,Baym1962,Atambo2025a,Atambo2025b}.
Finally, the dimer kernels are directly portable to first-principles codes:
the same two-kernel comparison can be implemented in existing BSE
workflows~\cite{Marini2009} to test whether the regime crossing survives in
real materials.

Two methodological limitations of the present static benchmark must be explicitly acknowledged. First, the kernel comparison relies on the Tamm-Dancoff approximation (TDA), which neglects resonant-antiresonant coupling. For the dimer, these coupling terms are of order $\mathcal{O}(K^2/E_{\rm gap})$. Given the large quasiparticle gap relative to the kernel elements, the TDA shifts the excitation energies by less than $0.01\,t$, and we have verified that the full BSE matrix yields a regime crossing identical to the TDA within numerical precision. Second, the present work compares static kernels against exact spectra but does not provide a baseline run with a frequency-dependent kernel. The deterioration of both static constructions beyond $V \simeq 2\,t$ is attributed to missing retardation physics, but explicitly demonstrating that a dynamical BSE kernel recovers the deep-binding branch remains a necessary extension of this work.

The broader methodological conclusion is that benchmarking an approximation
against exact answers is necessary but insufficient; the scientifically
decisive quantity is the structure of the error, and in particular whether a
correct number arises from a controlled expansion or from a cancellation. For
the $GW$+BSE pipeline, the answer is regime-dependent, and the regime is
knowable.

\begin{acknowledgments}
The author acknowledges Kenya Education Network (KENET) research services for
computing resources.
\end{acknowledgments}

\appendix
\section{Continuous tracking of the $GW$ quasiparticle poles}
\label{app:poles}

To ensure continuous, artifact-free trajectories of the quasiparticle poles as a function of the interaction parameters, we avoid grid-based root finding and polynomial branch-sorting. Instead, the physical quasiparticle poles are tracked continuously by minimizing the smallest singular value of the inverse dressed propagator,
\begin{equation}
\sigma_{\min}(\omega) = \min \text{svd}\left[ \omega I - h - \Sigma^{\rm static} - \Sigma^c(\omega) \right].
\end{equation}
By seeding the minimization at each parameter step with the pole location from the previous step, we enforce continuity across avoided crossings and guarantee that the tracked trajectories correspond strictly to the physical quasiparticle branches rather than spurious roots or satellite poles.

\section{Explicit evaluation of the vertex term $iG(\delta W/\delta G)$}
\label{app:vertex}

Continuum arguments suggest that the vertex term $iG(\delta W/\delta G)$ exactly cancels the screened exchange from the direct derivative, restoring the bare Coulomb exchange $v$. To test this explicitly in the dimer, we evaluate the functional derivative by propagating a coherence-direction perturbation of the reference Hamiltonian, $C = \psi_o\psi_u^T + \psi_u\psi_o^T$, through the full RPA chain to $W$. The static polarizability takes the rank-one form $\chi^0(0) = -\gamma \rho\rho^T$ with $\gamma=2/\Delta\varepsilon$ and $\rho_i = \psi_o(i)\psi_u(i)$. By the Sherman-Morrison formula, the statically screened interaction is
\begin{equation}
W(0) = v - \frac{\gamma a a^T}{1 + \gamma s},
\end{equation}
where $a = v\rho$ and $s = \rho^T v\rho$. The screened exchange is $\rho^T W \rho = s/(1+\gamma s)$. 

Varying $W$ through the $G$-dependence of $\rho$ and $\Delta\varepsilon$ yields the vertex contribution in the exchange channel. Closing the vertex with the static occupied Green's function line, we compute the exchange projection of the vertex term. If the continuum identity held, this would exactly equal the target $v_{ex} - W_{ex}$. Table~\ref{tab:vertex} reports the numerical evaluation of the vertex term, the target value, and the residual across the parameter space. The residual is of order $t$, not a rounding-level remainder. The vertex cancellation is therefore broken in the finite, static, rank-one polarizability of the dimer. 

\begin{table}[h]
\caption{Evaluation of the vertex term $iG(\delta W/\delta G)$ in the exchange channel for representative parameter points (units of $t$). The residual is the difference between the computed vertex term and the target $v_{ex} - W_{ex}$ required by the continuum cancellation identity.}
\label{tab:vertex}
\begin{ruledtabular}
\begin{tabular}{ccccc}
$\Delta/t$ & $V/t$ & Vertex term & Target ($v_{ex}-W_{ex}$) & Residual \\
\hline
0.50 & 0.0 & +3.771 & +1.216 & +2.554 \\
0.50 & 1.5 & +2.566 & +0.627 & +1.939 \\
1.00 & 2.5 & +1.525 & +0.210 & +1.315 \\
0.20 & 0.0 & +3.591 & +1.314 & +2.277
\end{tabular}
\end{ruledtabular}
\end{table}

\clearpage

\section*{Figure captions}

\begin{figure}[b]
\includegraphics[width=\columnwidth]{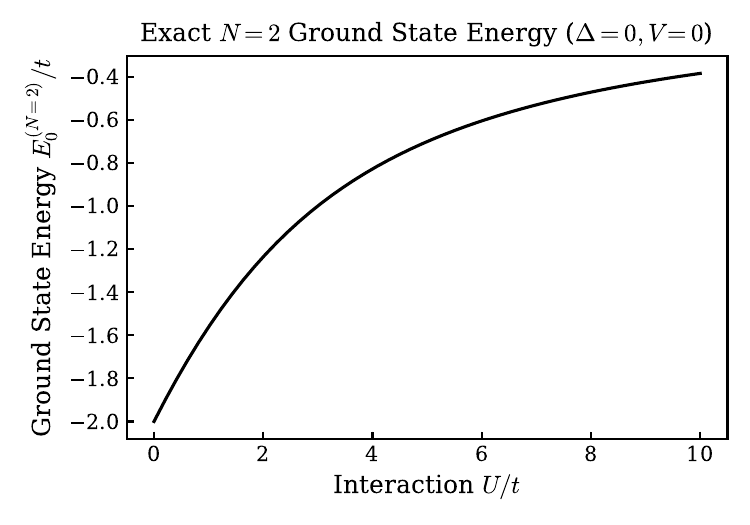}
\caption{Exact $N=2$ ground-state energy of the dimer as a function of $U/t$ at $\Delta=0$ and $V=0$, interpolating between the non-interacting limit $E_0=-2t$ and the atomic limit, and matching the analytic singlet expression $E_0=U/2-\sqrt{U^2+16t^2}/2$.}
\label{fig:gs}
\end{figure}

\begin{figure*}[t]
\includegraphics[width=\textwidth]{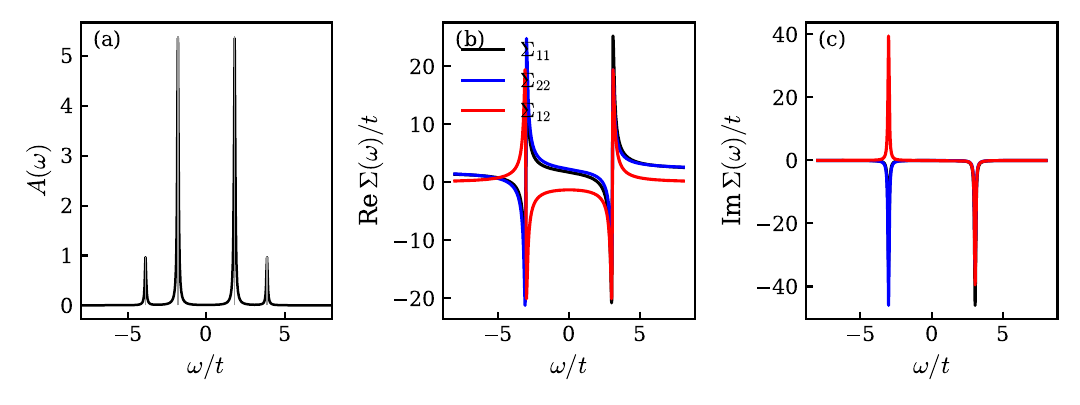}
\caption{(a) Exact spectral function $A(\omega)$ per spin at
$(U,V,\Delta)=(4,0,0.5)$, with sticks at the exact Lehmann poles.
(b), (c) Real and imaginary parts of the exact self-energy
$\Sigma_{\rm exact}(\omega)$ obtained by Dyson inversion of the exact Green's
function against the consistent non-interacting reference.}
\label{fig:sigma}
\end{figure*}

\begin{figure*}[t]
\includegraphics[width=\textwidth]{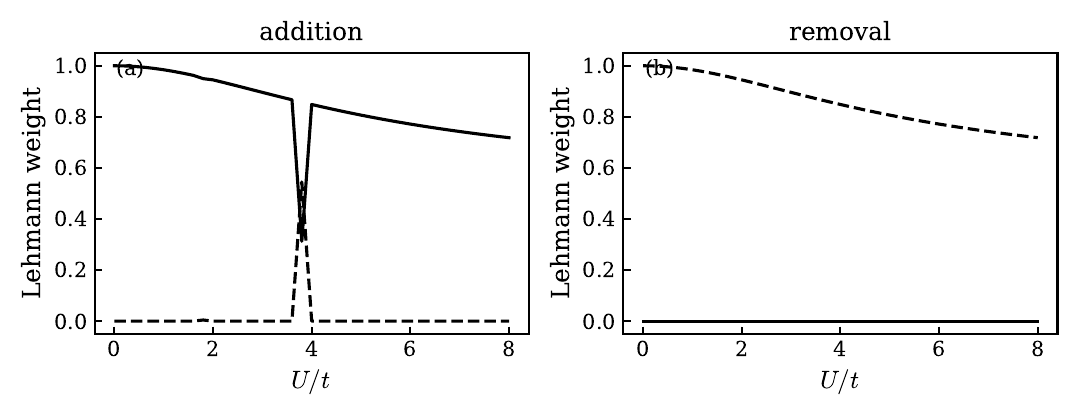}
\caption{Lehmann weights of the addition and removal poles as a function of
$U/t$. The branch whose weight flows to unity as $U\rightarrow0$ defines the
quasiparticle pole; the branch whose weight vanishes defines the
correlation-induced satellite.}
\label{fig:weights}
\end{figure*}

\begin{figure*}[t]
\includegraphics[width=\textwidth]{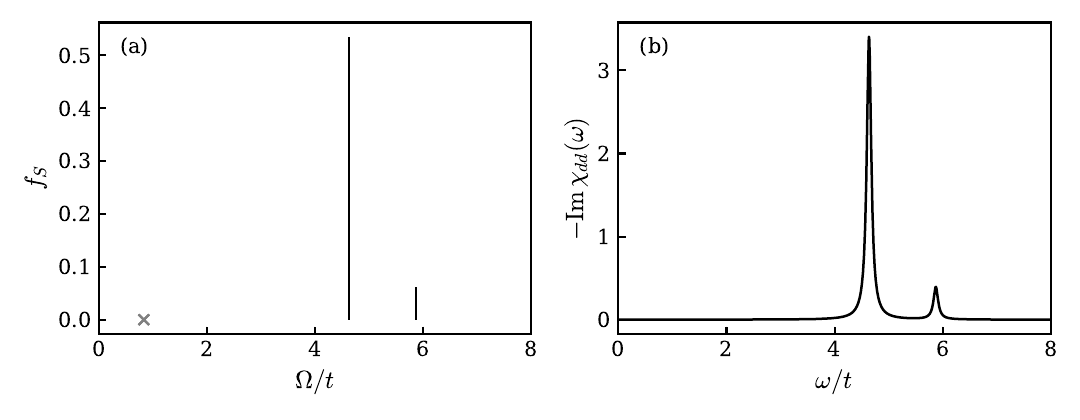}
\caption{(a) Exact neutral spectrum: dipole-active singlets (sticks) and dark
triplets (crosses). (b) Dipole loss function $-{\rm Im}\,\chi_{dd}(\omega)$.
The static-polarizability sum rule, Eq.~\eqref{eq:sumrule}, is satisfied to six digits.}
\label{fig:neutral}
\end{figure*}

\begin{figure*}[t]
\includegraphics[width=\textwidth]{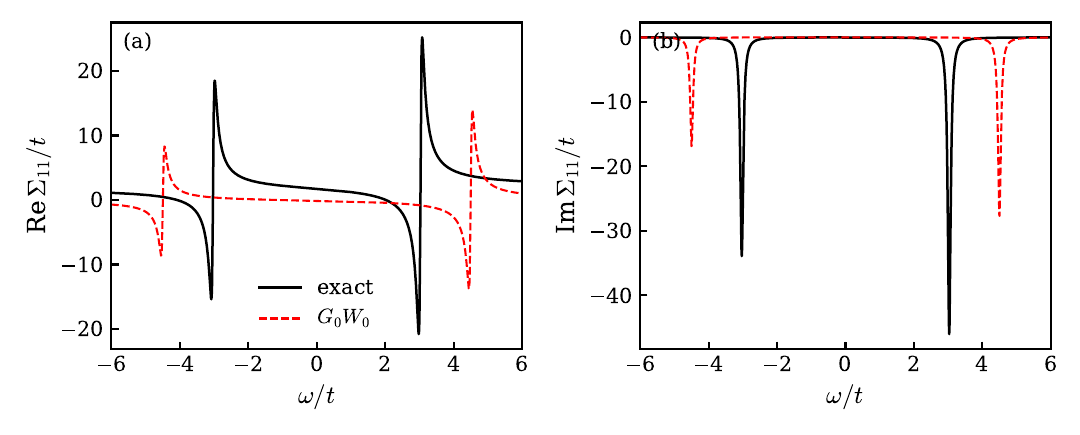}
\caption{Exact self-energy (solid) versus $G_0W_0$ self-energy (dashed). $GW$
reproduces the qualitative pole structure but misplaces the quasiparticle
poles by $0.44\,t$ and the satellites by $0.63\,t$.}
\label{fig:gw}
\end{figure*}

\begin{figure}[t]
\includegraphics[width=\columnwidth]{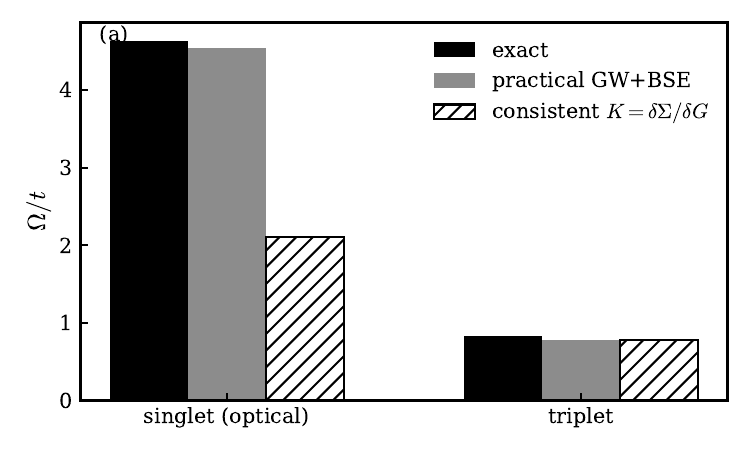}
\caption{Neutral excitation energies at $(U,V,\Delta)=(4,0,0.5)$: exact,
practical $GW$+BSE, and frozen-$W$ derivative $K=\delta\Sigma^{GW}/\delta G|_W$. The
practical construction reproduces the singlet through accidental
cancellation; the frozen-$W$ construction fails in the singlet channel while
both succeed in the exchange-free triplet channel.}
\label{fig:bars}
\end{figure}

\begin{figure*}[t]
\includegraphics[width=\textwidth]{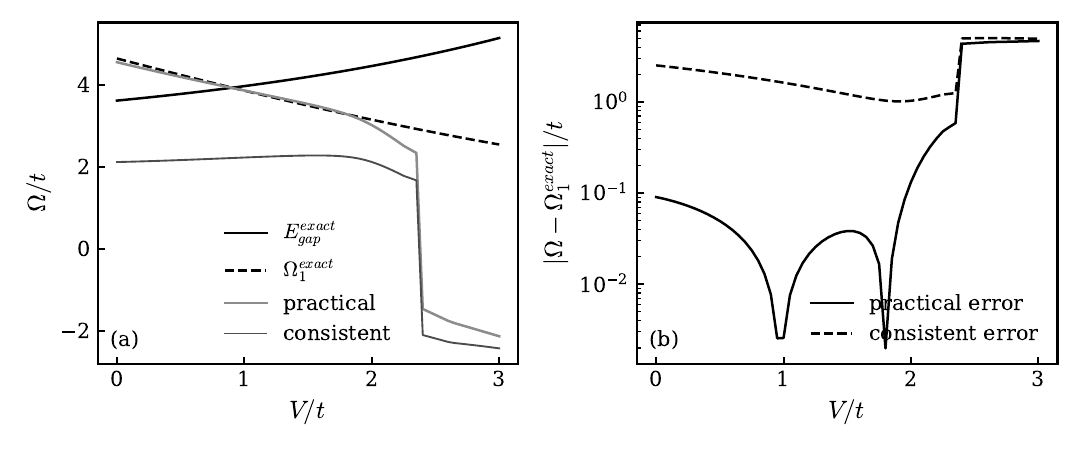}
\caption{(a) Exact quasiparticle gap and optical gap versus $V/t$, together
with the practical and frozen-$W$ BSE singlets evaluated with the analytic
pole solver; the bound exciton forms for $V\gtrsim1.0\,t$. (b) Excitation
errors on a logarithmic scale, showing cancellation sweet spots near
$V\simeq0.97\,t$ and $V\simeq1.80\,t$ and the systematic failure of the
frozen-$W$ derivative kernel; beyond $V\simeq2\,t$ both static constructions
deteriorate.}
\label{fig:crossing}
\end{figure*}

\end{document}